\documentstyle[a4,12pt,amssymb,times]{article}

\newcommand{\be}{\begin{equation}}
\newcommand{\ee}{\end{equation}}
\newcommand{\bea}{\begin{eqnarray}}
\newcommand{\eea}{\end{eqnarray}}

\def\p1{\pi_1}

\def\l{\lambda}
\def\f{\phi}

\def\r{\rho}

\def\pmas{\partial_+}
\def\pmen{\partial_-}

\renewcommand{\thefootnote}{\fnsymbol{footnote}}

\begin{document}

\begin{titlepage}

\vspace*{\stretch{0}}

\begin{center}
{\Large\bf Low-energy scattering of extremal black holes by neutral
matter}
\\[0.5cm]
A. Fabbri$^{\rm a}$\footnote{\tt fabbria@bo.infn.it},
D. J. Navarro$^{\rm b}$\footnote{\tt dnavarro@ific.uv.es} and
J. Navarro-Salas$^{\rm b}$\footnote{\tt jnavarro@ific.uv.es}
\\[0.5cm]
{\footnotesize
a) Dipartimento di Fisica dell'Universit\`a di Bologna and INFN
sezione di Bologna,\\
Via Irnerio 46, 40126 Bologna, Italy.
\\[0.5cm]
b) Departamento de F\'{\i}sica Te\'orica and IFIC, Centro Mixto
Universidad
de Valencia-CSIC.\\
Facultad de F\'{\i}sica, Universidad de Valencia, Burjassot-46100,
Valencia,
Spain.}
\end{center}

\bigskip

\begin{abstract}

We investigate the decay of a spherically symmetric near-extremal
charged black hole, including back-reaction effects, in the
near-horizon region. The non-locality of the effective action
controlling this process allows and also forces us to introduce a
complementary set of boundary conditions which permit to determine
the asymptotic late time Hawking flux. The evaporation rate goes
down exponentially and admits an infinite series expansion in
Planck's constant. At leading order it is proportional to the
total mass and the higher order terms involve higher order momenta
of the classical stress-tensor. Moreover we use this late time
behaviour to go beyond the near-horizon approximation and comment
on the implications for the information loss paradox.

\end{abstract}

\vspace*{\stretch{1}}

\begin{flushleft}
PACS number(s): 04.70.Dy, 04.62.+v
\end{flushleft}

\end{titlepage}

\newpage

\renewcommand{\thefootnote}{\arabic{footnote}}
\setcounter{footnote}{0}

\section{Introduction}

The discovery that black holes emit thermal radiation \cite{h} has
been considered a sign that the evaporation process implies a loss
of quantum coherence \cite{h2}. However, it has been stressed
\cite{th} that gravitational back-reaction effects could change
the standard picture of black hole decay. In particular, 't Hooft
\cite{th} suggested that the interaction between the infalling
matter and the outgoing radiation could preserve the unitarity of
the process through non-local effects.\\

One can consider a simplified scenario analysing the scattering of
an extremal Reissner-N\"ordstrom (RN) black hole by low-energy
massless neutral particles. The existence of a stable ground state
(extremal configuration, with vanishing Hawking temperature $T_H$)
avoids to encounter the problem of the singularity at the large
stages of the evaporation (indeed, in the case of the
Schwarzschild black holes $T_H \sim 1/M$ grows without bound) . It
is this feature which makes the process more tractable than the
evaporation of uncharged black holes. Moreover, restricting the
problem to spherically symmetric configurations, one can maintain
the main physical ingredients of the problem while at the same
time simplifying the mathematics involved. The resulting model,
with back-reaction effects included, was studied by Strominger and
Trivedi \cite{st} in the adiabatic approximation, and numerically
by Lowe and O'Loughlin \cite{lo} (see also \cite{sp}).\\

If we also restrict the analysis to a region very close to the
horizon we can describe the physical process by an effective
theory which turns out to be equivalent to a solvable
two-dimensional model. The effective model remains solvable also
at the one-loop quantum level and it has been studied in
\cite{fnn1,fnn2}. We shall summarize its main ingredients in
section 2. Since the effective action is non-local, a crucial
point to properly define the quantum theory is to select the
appropriate boundary conditions associated to the non-local terms
of the semiclassical equations of motion. In references
\cite{fnn1,fnn2} we chose the boundary conditions in such a way
that they naturally describe the evaporation of the black hole
from the point of view of an infalling observer very close to the
horizon (see also \cite{dl}). In section 3 we shall consider an
alternative set of boundary conditions which turns out to be very
relevant from the physical point of view because it corresponds to
an asymptotic observer at late retarded times. This is just the
part of future null infinity which can still be described by our
model. These two sets of boundary conditions are not compatible
(up to the extremal, static configuration). This fact can be
connected with the principle of complementarity
\cite{th,svv,stu,s2}, which states that the simultaneous
measurements made by an external observer and those made by an
infalling observer crossing the horizon are forbidden. The
solution we get for the new boundary conditions will be given in
section 4 and it is very different in form from the original one
(they indeed provide two different descriptions of the evaporation
process). However we will crucially impose that they match at the
end-point of the evaporation since then both solutions become
extremal. This matching condition allows us to determine the
asymptotic late time Hawking flux, including back-reaction
effects. In contrast with the standard picture, the Hawking flux
goes down exponentially at late times and it is not proportional
to the total mass of the classical incoming matter. Instead, we
find that it is proportional to a parameter which admits an
infinite series expansion in Planck's constant. At leading order
this parameter is the total mass and the higher order terms
involve higher order momenta of the classical stress-tensor. One
can go beyond the near-horizon approximation to evaluate the
Hawking flux by requiring energy conservation. We shall do it in
section 5 for the simplest case obtained by perturbing the
extremal black hole by means of a shock wave. All these results
have, potentially, far reaching consequences for the information
loss problem and we shall comment on it in the final section.

\section{The near-horizon model}

Imposing spherical symmetry to the Einstein-Maxwell theory
\be
\label{4ds}
ds_{(4)}^2= d\bar{s}_{(2)}^2 + 4l^2\f d{\Omega}^2 \, ,
\ee
where $l^2$ is Newton's constant, the corresponding dimensional
reduction leads to
a two-dimensional theory. If one rescales the metric by
\be
ds_{(2)}^2 = \sqrt{\f} d\bar{s}_{(2)}^2 \, ,
\ee
the two-dimensional action turns out to be
\be
\label{2action}
I = \int dx^2 \sqrt{-g} \left( R\f + l^{-2} V(\f) \right) \, ,
\ee
where
\be
V(\f) = (4\f)^{-\frac{1}{2}} - q^2 (4\f)^{-\frac{3}{2}} \, .
\ee
The extremal black hole radius $r^2_0=4l^2\f_0$ is recovered when
$V(\f_0)=0$, and
expanding $\f$ around $\f_0=\frac{q^2}{4}$ ($\f=\f_0+\tilde\f$) the action
(\ref{2action}) leads
to the Jackiw-Teitelboim model \cite{JT}
\be
\label{jtaction}
I = \int d^2x \sqrt{-g} \left[ (R +
\frac{4}{l^2q^3}) \tilde{\f} - \frac{1}{2} |\nabla f|^2 \right] \, ,
\ee
where we have added a matter field $f$ representing a four-dimensional
spherically
symmetric scalar field which propagates freely in the region close to the
horizon.
To properly account for back-reaction effects we have to consider the
corresponding
one-loop effective theory. Therefore we have to correct (\ref{jtaction})
by adding
the Polyakov-Liouville term \cite{P}
\bea
\label{paction}
I &=& \int d^2x \sqrt{-g} \left(R \tilde{\phi} + 4
\lambda^2 \tilde{\phi} -\frac{1}{2} \sum_{i=1}^N |\nabla f_i|^2\right)
\nonumber \\
&-&  \frac{N\hbar}{96\pi} \int d^2x \sqrt{-g} R \; \square^{-1} R +
\frac{N\hbar}{12\pi} \int d^2x \sqrt{-g} \lambda^2 \, ,
\eea
where we have considered the presence of $N$ scalar fields to enforce
that the
above effective action captures the proper quantum theory in the large N
limit
(keeping $N\hbar$ constant). In this limit the fluctuations of the
gravity
degrees of freedom can be neglected \cite{s}. Note that the
Polyakov-Liouville
action has a cosmological constant term which has been fixed ($\lambda^2
=
l^{-2}q^{-3}$) to ensure that the extremal configuration remains a
solution of
the quantum theory. In conformal gauge $ds^2=-e^{2\r}dx^+dx^-$ the
equations
of motion derived from (\ref{paction}) are
\bea
\label{eq1}
2\pmas \pmen \r + \l^2 e^{2\r} &=& 0 \, , \\
\label{eq2}
\pmas \pmen \tilde{\f} + \l^2 \tilde{\f} e^{2\r} &=& 0 \, , \\
\label{eq3}
\pmas \pmen f_i &=& 0 \, , \\
\label{eq4}
-2\partial^2_{\pm} \tilde{\f} + 4 \partial_{\pm} \rho \partial_{\pm}
\tilde{\f} &=& T^f_{\pm \pm} - \frac{N\hbar}{12\pi} t_{\pm} - \\
&& \frac{N\hbar}{12\pi} \left( (\partial_{\pm} \r )^2 -
\partial_{\pm}^2 \r \right) \, , \nonumber
\eea
where the chiral functions $t_{\pm}(x^{\pm})$, coming from the
non-locality of
the Polyakov-Liouville action, are related with the boundary conditions
of
the theory associated with the corresponding observers. The equation
(\ref{eq1}) is the Liouville equation with a negative cosmological
constant.
It has a unique solution up to conformal coordinate transformations. It
is
very convenient to choose the following form of the metric
\be
\label{lmetric}
ds^2 = -\frac{2l^2q^3 dx^+ dx^- }{(x^{-}-x^{+})^2} \, ,
\ee
which, in turn, is a way to fix the conformal coordinates
$x^{\pm}$, up to M\"obius transformations. In these coordinates
only the $t_{\pm}$ terms survive in the quantum part of the
constraints (\ref{eq4}), i.e. the semiclassical stress tensor is
just
\be
\label{sets}
\left< T_{\pm\pm}\right> =-\frac{N\hbar}{12\pi}t_{\pm}\> ,
\ee
and the relevant information
of the solutions is therefore encoded in the field $\tilde{\f}$.\\

In the gauge defined by the metric (\ref{lmetric}) the solution to the
equations
of motion is
\be
\tilde{\f} = \frac{1}{2} \pmas F(x^+) + \frac{F(x^+)}{x^--x^+} +
\frac{1}{2} \partial_- G(x^-) + \frac{G(x^-)}{x^+-x^-} \, , \ee
where the chiral functions $F(x^+)$, $G(x^-)$ are related to the
boundary functions $t_{\pm}(x^{\pm})$ \bea - \pmas^3 F &=&
-\frac{N\hbar}{12 \pi} t_+(x^+) + T^f_{++} \, , \\ - \pmen^3 G &=&
-\frac{N\hbar}{12 \pi} t_-(x^-) \, . \eea The crucial point  is
then to choose the suitable functions $t_{\pm}(x^{\pm})$.

\section{Boundary conditions}

The choice of the functions $t_{\pm}(x^{\pm})$ should be done on the
basis of physical considerations. The extremal black hole can be
described by the solution (up to M\"obius transformations)
\be
\tilde{\f} = \frac{lq^3}{x^--x^+} \, ,
\ee
where the coordinates $x^-$, $x^+$ can be identified with the classical
Eddington-Finkelstein coordinates $u$, $v$.
To match the extremal solution with a
near-extremal one necessarily requires the vanishing of $\partial_-^3 G$ and
therefore
\be
\label{bcnh1}
t_-(x^-) = 0 \, ,
\ee
thus implying that
\be
\label{outflux}
\langle T_{--} \rangle = 0 \, .
\ee

The point now is to choose the function $t_+(x^+)$. Due to
(\ref{outflux}) we can
also write a generic metric obeying the equations of motion in the
ingoing Vaidya-type gauge
\be
\label{mvaidya}
ds^2=-\left( \frac{2\tilde x^2}{l^2q^3} - l\tilde{m}(v) \right)
dv^2 + 2dvd\tilde x \, ,
\ee
where $\tilde{x}=l\tilde{\f}$ and
\be
\partial_v \tilde{m}(v) = T^f_{vv} - \frac{N\hbar}{12\pi} t_v(v) \, .
\ee
If the incoming classical matter $T^f_{vv}$ starts at $v_i$ and is
turned off at
some advanced time $v_f$ we have
\be
ds^2 = -\frac{2\tilde{x}^2}{l^2q^3} dv^2 + 2dvd\tilde{x} \, ,
\ee
before $v_i$. This solution can be brought into the the form
(\ref{lmetric}) with
the coordinate change
\bea
\label{vvacuum}
x^+ &=& v \, , \\
x^- &=& v + \frac{l^2q^3}{\tilde{x}} \, .
\eea
However, for $v>v_f$ the analysis is more involved. Let us first
simplify the
problem and consider that (\ref{mvaidya}) is the classical solution.
Therefore we
have (for $v>v_f$)
\be
\label{clmetric}
ds^2=-\left( \frac{2\tilde x^2}{l^2q^3} - l\tilde{m_{cl}}(v_f) \right)
dv^2 + 2dvd\tilde x \, .
\ee
This solution can also be transformed into (\ref{lmetric}) with the
coordinate
change
\bea
\label{vevaporating}
v &=& x^+_0 + \sqrt{\frac{2lq^3}{\tilde{m_{cl}}(v_f)}} {\mathrm arctanh}
\sqrt{\frac{\tilde{m_{cl}}(v_f)}{2lq^3}} (x^+-x^+_0) \, , \\
\tilde{x} &=& lq^3 \frac{1 -
\frac{\tilde{m_{cl}}(v_f)}{2lq^3}(x^+-x^+_0)
(x^--x^+_0)}{x^--x^+} \, ,
\eea
where $x^+_0$ is an integration constant.\\

Since the incoming classical and quantum fluxes vanish before $v_i$ we
should have
\be
t_v(v) = 0 \, ,
\ee
and therefore, according to (\ref{vvacuum}) and (\ref{vevaporating}) and
the
transformation law for the t's functions
\be
t_+(x^+) = \left( \frac{dv}{dx^+} \right)^2 t_v(v) + \frac{1}{2} \{ v,
x^+ \} \, ,
\ee
we get
\be
\label{tclass}
t_+(x^+) = \frac{2lq^3}{\tilde{m}_{cl}(v_f)} \frac{1}{\left(
\frac{2lq^3}{\tilde{m}_{cl}(v_f)} - (x^+-x^+_0)^2 \right)^2} \, ,
\ee
for $x^+>x^+_f$.\\

The above boundary condition has the following drawbacks

\begin{itemize}

\item It has been calculated according to the classical solution
(\ref{clmetric}).
So the back-reaction effects have not been included.

\item It requires that $x^+>x^+_f$ and it is unclear how to match with
the
condition $t_+(x^+)=0$ for $x^+<x^+_i$.

\end{itemize}

We can solve these problems just considering
\be
\label{bcnh2} t_+(x^+) = \frac{1}{2} \{ v, x^+ \} \, , \ee where
the Eddington-Finkelstein type coordinate $v$ is the one appearing
in the evaporating metric (\ref{mvaidya}). It is worth remarking
that the relation $x^+=x^+(v,\hbar)$ is no longer given by the
classical expression (\ref{vevaporating}), but rather it will be
determined once we solve the semiclassical equations of motion.
Therefore (\ref{bcnh2}) incorporates the back-reaction effects in
a self-consistent way, in contrast with the choice
(\ref{tclass}).\\

The above discussion may appear rather surprising since the equation
(\ref{outflux}) means that
there is not Hawking radiation at all. The evaporation is due to the
negative
incoming flux given by
\be
\langle T_{++} \rangle = -\frac{N\hbar}{24\pi} \{ v, x^+ \} \, ,
\ee
as measured by a free falling observer. However for an outside
observer the black hole shrinks due to the Hawking radiation. In
fact, with a fixed classical background, it is given by the
expression
\be
\langle T_{uu} \rangle = -\frac{N\hbar}{24\pi} \{ u_{in}, u \} \, ,
\ee
where $u_{in}$ is the outgoing null coordinate of the extremal solution.
But we know
that $u_{in}=x^-$ and this implies that the Hawking flux is proportional
to the
classical incoming mass $m_{cl}(v_f)$ (see the appendix)
\be
\label{lateflux}
\langle T_{uu} \rangle = \frac{N\hbar}{24\pi
lq^3} m_{cl}(v_f) \, .
\ee
This corresponds to the constant
thermal flux of near-extremal Reissner-N\"ordstrom black holes
measured by the asymptotic observer at future null infinity at
late times. Moreover (\ref{lateflux}) also reflects the fact that
the late time behaviour of the Hawking radiation depends only on
the total classical mass of the matter forming the near-extremal
black hole. There is not dependence on the details of the incoming
matter.\\

Now we have arrived at an apparent contradiction. The quantum
equations, which incorporate back-reaction effects, imply that
$\langle T_{--} \rangle = 0$, but our last argument shows that we
have indeed Hawking radiation. This puzzle is solved by invoking
the principle of complementarity \cite{th}. According to it we
cannot have a detailed description of the physics given by an
infalling observer and, simultaneously, by an asymptotic one.
Therefore, with this idea and the above discussion in mind, it
seems natural to consider the following boundary condition
\be
\label{bci1} t_+(x^+) = 0 \, ,
\ee
meaning that for the outside
observer ($v>>v_f$) there is not incoming quantum flux. This
boundary condition allows us to introduce a generic metric
satisfying the equations of motion in the outgoing Vaidya-type
gauge
\be
\label{mvaidyaout}
ds^2=-(\frac{2\tilde x^2}{l^2q^3}-l\tilde{m}(u))du^2 - 2dud\tilde x \, .
\ee
Then for the function $t_-(x^-)$ we have to choose
\be
\label{bci2}
t_-(x^-) = -\frac{1}{2} \{ u, x^- \} \, ,
\ee
since it reproduces the Hawking-type flux
\be
\label{transformed}
\langle T_{uu} \rangle = \left( \frac{dx^-}{du} \right)^2 \langle
T_{--}
\rangle = \frac{N\hbar}{24\pi} \left( \frac{dx^-}{du} \right)^2 \{ u,
x^- \} =
-\frac{N\hbar}{24\pi} \{ x^-, u \} \, .
\ee
As before, the relation $x^-=x^-(u,\hbar)$ is dynamical and it can only be
determined once we solve the
complete set of equations. In the limit $\hbar \rightarrow 0$ we
reproduce the
coordinate change obtained from the classical solutions, but in general
we will have
an infinite series expansion in $\hbar$ expressing the large quantum
effects of
back-reaction.\\

We want to finish our discussion on the boundary conditions by
stressing again that this alternative sets of conditions fits with
the idea of complementarity. The conditions (\ref{bcnh1}),
(\ref{bcnh2}) are the natural ones to describe the evaporation
process for an infalling observer very close to the horizon and
correspond to a negative influx of radiation crossing the apparent
horizon and no outgoing flux. Alternatively, one can provide a
description of the evaporation process from the point of view of
an outside observer. The conditions (\ref{bci1}), (\ref{bci2})
give a positive outflux of radiation and vanishing incoming flux.
It is worth to remark the important fact that we cannot impose
simultaneously these conditions. Obviously, there is an exception
and it corresponds to the solution with $t_+(x^+)=0=t_-(x^-)$, but
it is just the extremal configuration.\\

In summary, our scheme excludes the fact of having simultaneously
Hawking radiation and an ingoing quantum flux. The Hawking
radiation does exist in the boundary conditions (\ref{bci1}),
(\ref{bci2}), although there one does not see the negative ingoing
flux. On the other hand, the infalling observer does not see
outgoing radiation and the evaporation is due to the ingoing
radiation.

\section{Solutions and Hawking radiation}

The suitable boundary conditions for the infalling observer
\bea
t_+(x^+) &=& \frac{1}{2} \{ v, x^+ \} \, , \\
t_-(x^-) &=& 0 \, ,
\eea
imply that the solution can be written as
\be
\label{vq}
\tilde\phi=\frac{F(x^+)}{x^- - x^+}+\frac{1}{2}F'(x^+) \, ,
\ee
where the function $F(x^+)$ satisfies the differential equation
\be
\label{nby}
F'''=\frac{N\hbar}{24\pi}\left( -\frac{F''}{F}
+\frac{1}{2}(\frac{F'}{F})^2\right) -T_{++}^f (x^+) \, .
\ee
The function $F(x^+)$ relates the coordinates $x^+$ and $v$
\be
\frac{dv}{dx^+}=\frac{lq^3}{F} \, , \ee and in terms of the mass
function $\tilde{m}(v)$ the differential equation for $F$ turns
out to be
\be
\label{mfv}
\partial_v \tilde{m}(v)=-\frac{N\hbar}{24\pi lq^3}\tilde{m}(v)
+T_{vv}^f(v)
\, .
\ee
If the incoming classical matter is turned off at some advanced time
$v_f$
then the evaporating solution approaches asymptotically the extremal
configuration  (up to exponentially small corrections) \cite{fnn1,fnn2}
\be
\label{remnant}
\tilde{\f} = \frac{F''(x^+_{{\mathrm int}})}{2} \frac{(x^+-x_{{\mathrm
int}}^-)(x^--x_{{\mathrm int}}^-)}{x^--x^+} \, ,
\ee
where ($x^{\pm}_{{\mathrm int}}$) represent the end-point coordinates
that
belong to the AdS$_2$ boundary ($x^+_{{\mathrm int}}=x^-_{{\mathrm
int}}$).\\

In the alternative description of the evaporation process,
suitable for the outside observer, the boundary conditions are
\bea
t_+(x+) &=& 0 \, , \\
t_-(x-) &=& -\frac{1}{2} \{ u, x^- \} \, .
\eea
The solution can then be written as
\be
\label{vv}
\tilde\phi= \frac{G(x^-)}{x^+ - x^-} +\frac{1}{2}G'(x^-) \, ,
\ee
where the function $G(x^-)$ verifies the differential equation
\be
\label{nbz}
G'''=-\frac{N\hbar}{24\pi}\left(
-\frac{G''}{G}+\frac{1}{2}(\frac{G'}{G})^2 \right) \, ,
\ee
and serves to relate the coordinates $x^-$ and $u$
\be
\frac{du}{dx^-}=-\frac{lq^3}{G(x^-)} \, .
\ee
The evaporating mass function $\tilde{m}(u)$ obeys now the equation
\be
\partial_u \tilde{m}(u)= -\frac{N\hbar}{24\pi lq^3}\tilde{m}(u) \, ,
\ee
which implies that
\be
\tilde{m}(u) = \tilde{m}_0 \, e^{-\frac{N\hbar}{24\pi lq^3}u} \, .
\ee The point now is how to determine the integration constant
$\tilde{m}_0$, but this is related to the choice of the "initial"
conditions for the differential equation (\ref{nbz}). Since the
alternative pair of boundary conditions are compatible in the
extremal configuration we shall impose that the two solutions
(\ref{vq}), (\ref{vv}) match at the end-point
$(x^+_{int},x^-_{int})$, where both solutions approach the
extremal one. It is worth noting that once we move away from it
the corrections to eq.(\ref{remnant}) will of course be different
in the two cases and this agrees with the idea of complementarity.
Moreover, such a requirement is certainly nonlocal (and this
reminds the sort of nonlocal effects advocated by 't Hooft)
because it implies that the form of the function $G(x^-)$ for
$x^-<x^-_{int}$ (and therefore $\left< T_{uu}\right>$ for finite
$u$) depends on the precise form of the solution at the end-point
(where $\left< T_{uu}\right>=0$). Expanding $G(x^-)$ around
$x^-_{int}$ and imposing that (\ref{vv}) be exactly
(\ref{remnant}) for $x^-\to x^-_{{\mathrm int}}$ we obtain \bea
\label{czk} G(x^-_{{\mathrm int}}) &=& F(x^+_{{\mathrm int}}) = 0
\, , \\ \label{ccd} G'(x^-_{{\mathrm int}}) &=& F'(x^+_{{\mathrm
int}}) = 0 \, , \\ \label{cdu} G''(x^-_{int}) &=& -F''(x^+_{int})
< 0 \, . \eea So $F$ and $G$ are solutions of the differential
equations (\ref{nby}) and (\ref{nbz}), which in the region where
$T_{vv}^f=0$ differ just for an overall sign in their r.h.s.
Moreover both solutions have similar boundary conditions, again up
to a sign, in $F''(x^+_{{\mathrm int}})= -G''(x^-_{{\mathrm
int}})$ where $x^+_{{\mathrm int}}=x^-_{{\mathrm int}}$. Therefore
$G(x^-)$ is functionally equal to $-F(x^+)$ after exchanging $x^+$
with $x^-$. $F''(x^+_{{\mathrm int}})$ uniquely fixes
$\tilde{m}(v_f)$ and so (\ref{cdu}) implies that
$\tilde{m}_0=\tilde{m}(v_f) \, e^{\frac{N\hbar}{24\pi lq^3}v_f}$.
Therefore the Hawking flux is
\be
\label{cicco}
\left< T_{uu}(u)\right> =\frac{N\hbar}{24\pi lq^3}\tilde{m}(u)=
\frac{N\hbar}{24\pi lq^3}\tilde{m}(v_f)e^{-\frac{N\hbar}{24\pi
lq^3}(u-v_f)}
\, ,
\ee
where the explicit expression for $\tilde{m}(v_f)$ is given by the
formal
solution to the equation (\ref{mfv})
\bea
\label{mtil}
\tilde{m}(v_f) &=& \sum_{n=0}^{\infty}
(-\frac{N\hbar}{24\pi
lq^3})^n\int_{-\infty}^{v_f}dv_1\int_{-\infty}^{v_1}dv_2
\nonumber \\
& & ....\int_{-\infty}^{v_n}dv_{n+1} T_{vv}^f(v_{n+1}) \>.
\eea
It is important to point out the fact
that $\tilde{m}(v_f)$ depends on the details of the collapsing
matter through all the higher-order momenta of the classical
stress tensor. We observe that for $\hbar\to 0$ $\tilde{m}(v_f)$
is the total classical mass of the collapsing matter and
(\ref{cicco}) recovers the constant thermal value of a static
near-extremal black hole (\ref{lateflux}). So when back-reaction
effects are neglected we loose the information of the initial
state.

\section{Beyond the near-horizon approximation}

The solvability of the near-horizon model studied in the previous
sections has allowed us to work out the late time behaviour of the
Hawking flux. Of course, one would like to know $\langle T_{uu}
\rangle$ for every $u$, but this is out of the reach of our model.
However, if the incoming matter has the form of a spherical null
shell
\be
T_{uu}^f = \Delta m \delta (v-v_0) \, ,
\ee
general physical requirements for $\langle T_{uu} \rangle$ are so
strong
as to determine it completely. At leading order in $\hbar$ we impose
that (from
now on we set $l=1$)
\be
\label{c1} \langle T_{uu}^f \rangle = \langle T_{uu}^f
\rangle_{NBR} + {\cal O} (\hbar^2) \, , \ee where $\langle
T_{uu}^f \rangle_{NBR}$ is the Hawking flux computed in the
classical background (no backreaction) defined by the matching of
the extremal black hole of mass $q$ (for $v<v_0$)
\be
ds^2 = - \left( 1 - \frac{q}{r} \right)^2 du_{in} dv \, ,
\ee
and the near-extremal one of mass $q+\Delta m$ as $v>v_0$
\be
ds^2 = - \frac{(r-r_+)(r-r_-)}{r^2} dudv \, .
\ee
The relation between $u$ and $u_{in}$ is given by
\be
\frac{du}{du_{in}} = \frac{(r-q)^2}{(r-r_+)(r-r_- )} \, , \ee and
the Hawking flux without back-reaction $\langle T_{uu}^f
\rangle_{NBR}$ is
\be
\langle T_{uu}^f \rangle_{NBR} = -\frac{N\hbar}{24\pi} \{ u_{in},
u \} \, , \ee which turns out to be \bea \langle T_{uu}^f
\rangle_{NBR} (u,m,q) &=& \frac{N\hbar}{24\pi} [
\frac{(m-q)(r-r_+) (r-r_-)(r^2+rq-q^2)}{r^5(r-q)^2}\\ \nonumber
&+& \frac{1}{2} \frac{(m-q)^2(r+q)^2}{r^4(r-q)^2} ] \, , \eea
where \bea \label{f1} r_{\pm} &=& m \pm \sqrt{m^2-q^2} \, , \\
\label{f2} m &=& q + \Delta m \, , \eea and
\be
\label{f3}
\frac{v_0-u}{2} = r + \frac{1}{r_+-r_-} \left[ r_+^2 \ln \left|
\frac{r_- r_+}{r_+} \right|
- r_{-}^2 \ln \left| \frac{r_- r_-}{r_-} \right| \right] \, .
\ee
Moreover, energy conservation implies that
\be
\label{c2} \int_{-\infty}^{+\infty} \langle T_{uu} \rangle du =
\Delta m \, . \ee In addition, $\langle T_{uu} \rangle$ should
verify, as $u\to -\infty$,
\be
\label{c3} \langle T_{uu} \rangle \sim \langle T_{uu}
\rangle_{NBR} = \frac{N\hbar \Delta m}{3\pi |u|^3} \, , \ee
because at early times the back-reaction can be
ignored\footnote{See the essay \cite{fnn3} for a comparison of
this problem to the Planck problem of black body radiation.}.
These conditions are satisfied automatically if $\langle T_{uu}
\rangle (u)$ is just the r.h.s. of the differential equation
\be
-\frac{dm}{du} = \frac{N\hbar}{24\pi} \langle T^f_{uu}
\rangle_{NBR} (u, m(u), q) \, , \ee where $m=m(u)$ and
$r_{\pm}=r_{\pm}(u)$, $r=r(m)$ are given by expressions similar to
(\ref{f1})-(\ref{f3}), and $m(u)$ verifies the initial condition
\be
m(u=-\infty) = q + \Delta m \, . \ee It is easy to see that this
proposal for $\langle T_{uu} \rangle (u)$ fulfills the conditions
(\ref{c1}), (\ref{c2}), (\ref{c3}), and it is very difficult to
imagine an alternative solution. Moreover the late time behaviour
of $\langle T_{uu} \rangle$ is also of the form
\be
\langle T_{uu} \rangle \sim \frac{N\hbar}{24\pi q^3} \tilde{m}_0
e^{-\frac{N\hbar}{24\pi lq^3}u} \, , \ee as $u\to +\infty$, where
$\tilde{m}_0$ is an integration constant. It can be shown
\cite{fnno} numerically that $\tilde{m}_0$ agrees with the
expression obtained in section 4 ($\tilde{m}_0=\Delta m
e^{\frac{N\hbar}{24\pi lq^3}v_0}$) thus providing a
self-consistency test of our approach.

\section{Conclusions}

In this paper we have studied the near-horizon effective theory
controlling the decay of a near-extremal charged black hole. We
have focused on the delicate point of how to choose the
integration functions $t_{\pm}$ coming from the non-locality of
the Polyakov-Liouville action. We have stressed the fact that it
is not possible to choose, simultaneously, non-vanishing functions
$t_{\pm}(x^{\pm})$. Since they are proportional, in a particular
coordinate system $\{ x^{\pm} \}$, to the quantum fluxes this
seems to lead to inconsistencies. The vanishing of $t_-$ implies
the absence of Hawking radiation, and the black hole shrinks due
to the negative incoming quantum radiation produced by $t_+$. We
have interpreted this apparently disturbing situation in terms of
a complementarity between the physical descriptions given by an
infalling observer and by an asymptotic one. Our model, which
captures the (near-horizon) quantum back-reaction in the large $N$
limit, dictates that there is not an unique choice for $t_{\pm}$,
up to the extremal configuration, and it seems natural to choose
$t_-(x^-)=0$ and $t_+(x^+)=\frac{1}{2} \{ v, x^+ \}$ for the
infalling observer and $t_+(x^+)=0$ and $t_-(x^-)=-\frac{1}{2} \{
u, x^- \}$ for the outside observer.\\

We would like to stress that eq. (\ref{cicco}) is the first
calculation of the Hawking radiation flux for RN black holes at
late times, which takes into account consistently back-reaction
effects, and in the large $N$ limit it is exact. Our result is
highly non-trivial because two different expansions in $N\hbar$
are implicit in (\ref{cicco}), one being associated to the
exponential $e^{-\frac{N\hbar}{24\pi lq^3}(u-v_f)}$ and the other
inside $\tilde{m}(v_f)$, see (\ref{mtil}). While the first
expansion is of no surprise, the second one is completely
unexpected on physical grounds. Actually it implies that the
information carried by the classical incoming matter can be read
from the late time Hawking radiation, which  admits an infinite
series expansion in $N\hbar$ and where each term involves
different momenta of the classical stress tensor. This is in
contrast with the predictions based on fixed background
calculations. When the back-reaction is ignored the late-time
Hawking flux goes to the constant thermal value (\ref{lateflux}) .
To deepen our result we can mention that the relation between the
coordinates $u_{in}$ and $u_{out}$ before and after the classical
influx of matter $T^f_{vv}$ is given by ($u_{out}\rightarrow
+\infty$)
\be
\frac{du_{out}}{du_{in}} \sim u^2_{out} (A-Be^{-Cu_{out}}) \, ,
\ee where $A$, $B$ and $C$ are positive integration constants
depending on $m(v_f)$. This also implies that the radiation is
quite different from the standard late-time thermal radiation
coming from the relation ($u_{out}\rightarrow +\infty$)
\be
\frac{du_{out}}{du_{in}} \sim e^{2\pi T_H u_{out}} \, , \ee where
$T_H$ is the Hawking temperature (for similar results see also
\cite{liroso}).\\

Nevertheless we have to remark that this result does not necessarily
means that the
"quantum information", in addition to the classical one given by
$T^f_{vv}$, is also
encoded in the late time radiation. In the standard picture of black
hole evaporation
the "quantum information" is encoded in the correlation between outgoing
and incoming
particle-antiparticle pairs. The outgoing (Hawking) radiation is
uncorrelated and
therefore represents a mixed state. In our scheme we have either ingoing
or outgoing
radiation, according to the observer. So, this suggests that for the
asymptotic observer
the outgoing radiation can only be correlated with itself. This opens
the interesting
possibility, using the proposal of section 5, of studying
the correlation functions between the outgoing radiation at early and late
times to see
whether or not it corresponds to a pure state. This must be done
numerically \cite{fnno}.

\section*{Acknowledgements}

This research has been partially supported by the DGICYT, Spain.
D. J. Navarro acknowledges the Ministerio de Educaci\'on y Cultura for a FPI
fellowship. We thank D. Amati, R. Balbinot, V. Frolov and W. Israel for useful
discussions. J. N-S.
thanks the Department of Physics of Bologna University for hospitality
during the early stages of this work.

\section*{Appendix}

We can determine the Hawking flux without back-reaction by matching
static
solutions. If $T^f_{vv}=m_1 \delta (v-v_1)$ we have
\be
ds^2=-\left( \frac{2\tilde x^2}{l^2q^3} - l\tilde{m}_{cl}(v) \right)
dv^2 + 2dvd\tilde{x} \, ,
\ee
where $\tilde{m}_{cl}(v)=m_1 \Theta (v-v_1)$. In conformal gauge we get
\be
ds^2 = -\frac{2\tilde{x}}{l^2q^3} du_{in} dv \, ,
\ee
for $v<v_0$, and
\be
ds^2 = - \left( \frac{2\tilde{x}}{l^2q^3} - lm_1 \right) du_{out} dv \,
,
\ee
for $v>v_0$. The matching at $v=v_1$ implies that
\be
u_{in} = v_0 + \sqrt{\frac{2lq^3}{m_1}} {\mathrm cotanh}
\sqrt{\frac{m_1}{2lq^3}}
(u_{out} - v_1) \, .
\ee
The Hawking flux is given by
\be
\langle T_{u_{out} u_{out}} \rangle = -\frac{N\hbar}{24\pi} \{
u_{in}, u_{out} \} = \frac{N\hbar}{24\pi lq^3} m_1 \, . \ee In the
case of two shock waves $T^f_{vv}=m_1 \delta (v-v_1)+m_2 \delta
(v-v_2)$ one obtains \bea u_{in} &=& v_0 +
\sqrt{\frac{2lq^3}{m_1}} {\mathrm cotanh} [
\sqrt{\frac{m_1}{2lq^3}} (v_1-v_0) + \nonumber \\ & &{\mathrm
arctanh} \sqrt{\frac{m_1}{m_1+m_2}} {\mathrm tanh}
\sqrt{\frac{m_1+m_2}{2lq^3}} (u_{out}-v_2) ] \, , \eea and then
\be
\langle T_{u_{out} u_{out}} \rangle = \frac{N\hbar}{24\pi lq^3} (m_1+m_2)
\, .
\ee
The argument can be repeated so on for an arbitrary finite set of shock waves
$T^f_{vv}=\sum_{i=1} \delta (v-v_i)$ and the result is
\be
\langle T_{u u} \rangle = \frac{N\hbar}{24\pi lq^3}
\tilde{m}_{cl}(v_f) \, , \ee where $\tilde{m}_{cl}(v_f)=\sum_i
m_i$ is the total mass of the incoming matter and $v_f=v_N$. It is
interesting to remark that the above Hawking flux does not see the
details of the incoming matter . It is only sensitive to the total
incoming mass. The information carried out by the classical stress
tensor is lost if one neglects the back-reaction effects in the
Hawking flux.



\begin{thebibliography}{99}

\bibitem{h}
S. W. Hawking, {\it Nature} {\bf 248} (1974) 30; {\it Comm. Math.
Phys.} {\bf 43} (1975) 199.

\bibitem{h2}
S. W. Hawking, {\it Phys. Rev.} {\bf D14} (1976) 2460.

\bibitem{th}
G. 't Hooft, {\it Nucl. Phys.} {\bf B355} (1990) 138.

\bibitem{st}
A. Strominger and S. P. Trivedi, {\it Phys. Rev.} {\bf D48} (1993) 5778.

\bibitem{lo}
D. A. Lowe and M. O'Loughlin, {\it Phys Rev.} {\bf D48} (1993) 3735.

\bibitem{sp}
E. Sorkin and T. Piran, {\it Phys. Rev.} {\bf D63} (2001) 124024.

\bibitem{fnn1}
A. Fabbri, D. J. Navarro and J. Navarro-Salas, {\it Phys. Rev. Lett.}
{\bf 85} (2000) 2434.

\bibitem{fnn2}
A. Fabbri, D. J. Navarro and J. Navarro-Salas, {\it Nucl. Phys.}
{\bf B595} (2001) 381.

\bibitem{dl}
K. Diba and D. A. Lowe, {\it The stress-energy tensor in soluble models
of
spherically symmetric charged black hole evaporation}, hep-th/0107137.

\bibitem{svv}
K. Schoutens, E. Verlinde and H. Verlinde, {\it Phys. Rev. } {\bf D48}
(1993) 2670; {\it Quantum black hole evaporation}, hep-th/9401081.

\bibitem{stu}
L. Susskind, L. Thorlacius and J. Uglum, {\it Phys. Rev.} {\bf D48}
(1993) 3743.

\bibitem{s2}
L. Susskind, {\it Phys. Rev. Lett.} {\bf 71} (1993) 2367; L. Susskind
and L.
Thorlacius, {\it Phys. Rev.} {\bf D49} (1994) 966.

\bibitem{JT}
R. Jackiw, in {``Quantum Theory of Gravity''}, edited by S. M.
Christensen
(Hilger, Bristol, 1984), p. 403; C. Teitelboim, in op. cit., p. 327.

\bibitem{P}
A. M. Polyakov, {\it Phys. Lett. } {\bf B103} (1981) 207.

\bibitem{s}
A. Strominger, {\it Les Houches lectures on black holes}, hep-th/9501071.

\bibitem{fnn3}
A. Fabbri, D.J. Navarro and J. Navarro-Salas, {\it A Planck-like problem
for quantum charged black holes}, gr-qc/0105061 (to be published in
{\it Gen. Rel. Grav.}).

\bibitem{fnno}
A. Fabbri, D. J. Navarro, J. Navarro-Salas and G. Olmo, work in
progress.

\bibitem{liroso}
S. Liberati, T. Rothman and S. Sonego, {\it Phys. Rev. } {\bf D62}
(2000) 024005.

\end{thebibliography}
\end{document}